\documentclass[journal]{IEEEtran}
\usepackage{amsmath,amssymb,amsfonts,dsfont, amsthm}

\usepackage{bm}
\usepackage{algorithm, algorithmicx}
\usepackage{algpseudocode}
\usepackage{subcaption}
\def\BibTeX{{\rm B\kern-.05em{\sc i\kern-.025em b}\kern-.08em
		T\kern-.1667em\lower.7ex\hbox{E}\kern-.125emX}}
\usepackage{makecell}
\usepackage{breqn}
\usepackage{tikz}
\usepackage[english]{babel}
\usepackage{multirow}
\usepackage{comment}
\newtheorem{theorem}{Theorem}[section]
\newtheorem{lemma}[theorem]{Lemma}
\usepackage{upgreek}
\usepackage{textcomp}
\newcommand{\textapprox}{\raisebox{0.5ex}{\texttildelow}}

\setcellgapes{3pt}

\begin{document}

\title{Enhanced Multi-Target Tracking in Dynamic Environments: Distributed Flooding Control in the Random Finite Set Framework}

\author{Aidan~Blair, Amirali~Khodadadian~Gostar, Alireza~Bab-Hadiashar, Xiaodong Li and~Reza~Hoseinnezhad
\thanks{Submitted 20 November 2024.}
\thanks{All authors are with RMIT University, Melbourne, Victoria (email: aidan.blair@rmit.edu.au).}}

\maketitle
	
\begin{abstract}
\textit{Tracking multiple targets in dynamic environments using distributed sensor networks is a challenging problem for situational awareness in connected autonomous vehicles (CAVs). In such scenarios, the network of mobile sensors must coordinate their actions to accurately estimate the locations and trajectories of multiple targets, balancing limited computation and communication resources with multi-target tracking accuracy. Multi-sensor control methods can improve the performance of these networks by enabling efficient utilization of resources and enhancing the accuracy of the estimated target states. This paper proposes a novel multi-sensor control method that utilizes flooding-based communication to address this problem, ensuring distributed consensus of optimal sensor actions throughout the sensor network. Our method improves computational tractability and enables fully distributed control, ensuring the scalability and flexibility necessary for real-time CAV applications. Experimental results on several challenging multi-target tracking scenarios demonstrate that our approach significantly improves both multi-target tracking accuracy and computation time over competing methods.}
\end{abstract}


\section{Introduction}
\label{sec:Introduction}
This article addresses the problem of distributed multi-sensor control for multi-target tracking, a key challenge in enhancing situational awareness in connected autonomous vehicles (\text{CAVs}). In distributed sensor networks, each sensor node independently collects measurements of targets within its field of view (\text{FoV}) and executes a stochastic multi-target filter on board that produces a multi-object posterior. These posterior probabilities are then exchanged across the network and fused locally at each sensor node, allowing each sensor to have situational awareness of the larger environment. This capability is essential for CAV systems, where continuous real-time information fusion from sensors on multiple vehicles enables safe and efficient autonomous operation in complex and dynamic environments.

\begin{figure}[t]
	\centering
	\includegraphics[width=\columnwidth]{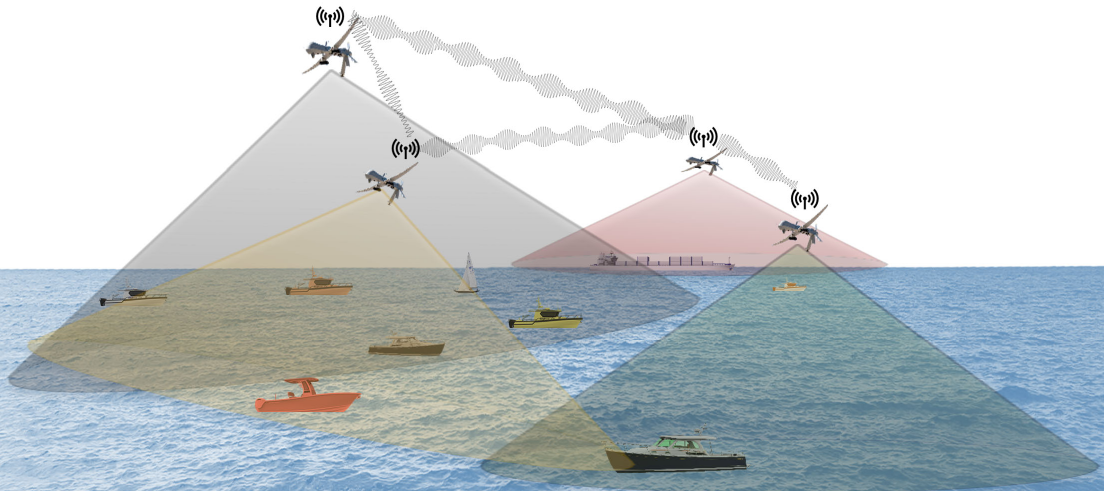}
	\caption{Multiple UAVs must cooperatively inspect part of the ocean for monitoring of marine vehicles on the sea surface.}
	\label{fig1}
\end{figure}

The need for multi-sensor control emerges when sensor states, such as position and orientation, can be adjusted via control commands, i.e. \textit{movement}. An instance illustrating the utilization of multiple sensors mounted on Unmanned Air Vehicles (UAVs) for inspecting the sea surface to detect and track sea vessels is depicted in Figure~\ref{fig1}. In this illustration, each UAV is required to autonomously determine its subsequent actions (in terms of transitioning to a new location and orientation), relying on the hypothesized future movements of other UAVs within the network. The decision must be made in a manner that optimizes the acquisition of \textit{most informative} measurements by the interconnected UAV network, aiming to obtain a situational awareness that is as comprehensive as possible.

Multi-sensor control is a complex and critical area of research that has gained significant attention in various fields, including but not limited to autonomous vehicle networks~\cite{ref:AutonomousVehicleControl}, surveillance and intelligence applications~\cite{ref:SurveillanceNetwork}, and defense applications of multi-target tracking~\cite{ref:MTTControl, ref:GroundAreaDetection}. In all such applications, a network of sensors are used to acquire and fuse information for situational awareness. Depending on how information is communicated in the network, it may be characterized as a centralized, decentralized, or distributed sensor network~\cite{ref:NetworkArchitecture,ref:DistributedArchitecture}. 

Centralized networks are made up of a central processing node that gathers data from every other node in the network, processes it, and distributes the results back to the other nodes. In decentralized networks, nodes form local clusters, with each cluster containing a central node that communicates with the other nodes in the cluster. In addition, these central nodes can communicate with other central nodes in the network, facilitating information transfer throughout the network. Distributed networks lack central processing nodes. Instead, nodes can only communicate with their nearest neighbors, usually within a limited communication range.  Distributed networks offer a pragmatic solution for implementing large-scale sensor networks, as the computational cost does not grow exponentially with the number of sensors as it does with centralized networks. In practical applications with large sensor networks, distributed communication and computing is the preferred option. Hence, we focus on distributed multi-sensor control.

Distributed networks are commonly divided into two categories with regard to inter-node communication: consensus and flooding. In consensus-based communication, the nodes in the network collaborate to reach an agreement or consensus on a certain value or state. This involves iterative exchanges of information among nodes until they converge to a common decision. Consensus algorithms aim to achieve agreement while minimizing information exchange and communication overhead~\cite{ref:xiao2005scheme, ref:ConsensusCPHD}. Flooding involves broadcasting information from a source node to all other nodes in the network. Each node that receives the broadcast information will then forward it to all of its neighbors. This process continues until all nodes in the network have received the information ~\cite{ref:flooding2017, ref:flooding2020}. 

Our proposed solution employs Random Finite Sets (RFS), a powerful Bayesian framework for multi-object tracking, wherein the sets of target states and observations are represented as random finite sets. Compared to other multi-target tracking frameworks such as particle filters and multiple hypothesis tracking~\cite{ref:JSCTrackingFusion}, the RFS framework has several advantages including implicitly handling dynamic numbers of targets and data association. A number of filters have been developed in this framework, including the well-known probability hypothesis density (PHD)~\cite{ref:MahlerBook} and multi-Bernoulli (MeMBer)~\cite{CBMEMBER_filter} filters. The latest development in RFS filters, called \textit{labeled RFS filters}, appends the label of each target into its single-target state and propagates target labels with their states to directly create \textit{target trajectories}~\cite{Vo2013,LMB_Vo2,reuter2015multiple}. RFS filters have found applications in various fields, including cell lineage tracking~\cite{ref:CellTracking}, intelligent transport systems (ITS)\cite{ref:NidaInteraction}, and information fusion\cite{Mahler2014AdvancesIS, ref:InformationFusionRFS}, especially in multi-target tracking applications.

Various sensor control methods have been proposed for Random Finite Sets (RFS) filters using different objective functions~\cite{ref:BrankoRistic,ref:MultiBernoulliControl,ref:VoControl, ref:HoangControl}. However, these methods have been explored mainly in single-sensor control scenarios. Recent works have introduced multi-sensor control for labeled RFS filters. For example, Jiang et al.~\cite{ref:MengMultiSensorControl} proposed a Cauchy-Schwarz divergence (CSD)-based objective function combined with Generalized Labeled Multi-Bernoulli (GLMB) filters. Another example is the multi-sensor control solution in which a task-driven cost (which is a mixture of the expected localization and cardinality errors) was chosen to be minimized~\cite{ref:Xiaoying}. In this work, a Labeled Multi-Bernoulli (LMB) filter was assumed to be running on board each sensor node. Panicker et~al.~\cite{PANICKER2020107451} developed a multi-sensor control solution that focuses on targets of interest, again with LMB filters running on each node. An information-theoretic objective function for RFS-based multi-sensor control was also proposed~\cite{ref:SabitaCauchySchwarz}.

This paper extends previous research into distributed multi-sensor control using distributed coordinate descent~\cite{ref:AidanICCAIS} by introducing the Distributed Flooding-based Sensor Control (DF-SC) algorithm, optimized for use with LMB filters across sensor nodes in a distributed network. DF-SC significantly reduces computational complexity while enhancing tracking accuracy and scalability for CAV applications. We also propose a constrained information-theoretic objective function tailored for the DF-SC algorithm, yielding improved performance over existing methods.

The structure of the paper is as follows: Section~\ref{sec:Related_Work} reviews current multi-sensor control methods with a focus on RFS filter-based approaches; Section~\ref{sec:Background} provides background information on the RFS framework and formulates the problem of multi-sensor control within the general distributed multi-target tracking application; Section~\ref{sec:Solution} details the proposed DF-SC algorithm; Section~\ref{sec:Experiment} presents experimental results, comparing our approach with state-of-the-art techniques; and Section~\ref{sec:Conclusion} concludes the paper.

    \section{Related Work}
	\label{sec:Related_Work}
	
	In the RFS framework, various solutions have been developed for multi-target tracking-related problems. Some examples include track-before-detect visual tracking applications~\cite{s20030929}, sensor management in target tracking applications~\cite{PANICKER2020107451,8455829}, and information fusion~\cite{9311857}. Several information fusion approaches have been explored for applications with LMB filters in place as tracking solution, ranging from Cooperative CSD-based fusion~\cite{ref:CauchySchwarzLMB} to Consensus-based fusion~\cite{ref:ConsensusLMB} and Complementary fusion~\cite{ref:Klupacs, ref:KlupacsFusion, ref:KlupacsComplementaryGCI}. 
	
	The communication between sensors in a distributed sensor network has been the subject of significant research, but distributed multi-sensor control has not received as much attention. Akselrod and Kirubarajan~\cite{ref:MarkovDecisionProcesses} proposed a distributed control algorithm for multi-target tracking by a swarm of UAVs, where measurements and associated tracks from each UAV are broadcast throughout the network and fused with the local measurements at each UAV. However, each UAV chooses an action to take without considering the actions of other UAVs, which could lead to multiple nearby UAVs converging on the same target, which is counterproductive in a large distributed network. To address this issue, Fu and Yang~\cite{ref:MultitargetTrackingMobileSensor} proposed a control framework aimed at maximizing tracking accuracy while guaranteeing tracking coverage. Each sensor can be allocated to track a particular target or group of targets, and sensor positions and allocations are shared distributively throughout the network to jointly optimize for tracking accuracy while ensuring that at least one sensor is allocated to every target.
	
	Yuan, Zhan, and Li~\cite{ref:DecentralizedQuadcopterControl} use a distinct approach by flocking quadcopter UAVs in a decentralized manner. Each UAV shares its local information with neighboring nodes, which is then used by a decentralized model predictive control (DMPC) flocking algorithm to form a quasi \(\alpha\)-lattice. In this application, sensors have access to information only from their neighboring sensors, limiting their ability to form a fully accurate lattice. Li et al.~\cite{ref:DistributedSparsityControl} recently proposed a fully distributed multi-sensor control approach that uses auctioned POMDPs to determine each sensor's action. While this approach is distributed and computationally efficient, it has the limitation of each possible action being taken by at most one sensor at a time.

    Coordinate Descent is a family of optimization techniques where each dimension/parameter is optimized sequentially while the others are kept fixed, approximately optimizing along coordinate directions. An overview of various coordinate descent algorithms is provided by Wright~\cite{ref:CoordinateDescentOverview}. Wang et al.~\cite{ref:Xiaoying} proposed a method for multi-sensor control of RFS filters using coordinate descent. Recently, a distributed coordinate descent-based method for multi-sensor control in multi-target tracking applications has been developed~\cite{ref:AidanICCAIS}. Genetic Algorithms (GAs) are stochastic population-based optimization methods, based on the principles of natural selection~\cite{ref:GeneticAlgorithms}. They do not require gradient information, are generic and more robust (i.e. less likely to get stuck in local optima), and can be applied to a wide range of optimization problems. Numerous control algorithms across many fields have used genetic algorithms for optimization, including UAV control~\cite{ref:GeneticAlgorithmUAV} and wireless sensor networks~\cite{ref:GeneticAlgorithmWSN}. Similar to GAs, Particle Swarm Optimization (PSO) based methods have been used to determine the optimal positions and velocities of sensors in multi-target tracking scenarios~\cite{ref:BearingsSensorManagement}.

    Generally, two types of objective functions have appeared in the sensor control literature: the \textit{task-driven} and the \textit{information-driven}. The former type is usually defined to directly optimize a particular aspect of tracking. For instance, in some applications, the highest priority is given to the \textit{coverage} aspect. Having a network of sensors with limited fields of view, distributing the sensors for maximum coverage could be achieved by choosing the \textit{cardinality estimate} returned by the fused pseudo-posterior. The resulting sensor control solution, originally designed for one sensor, is called ``Posterior Expected Number of Targets'' (PENT)~\cite{mahler2004probabilistic}. Alternatively, estimation of tracking error could be the particular aspect of the tracking task being optimized by sensor control. In that case, the objective function would be formulated as a cost function dependent on the expected error of estimation or tracking after a sensor action is taken~\cite{ref:PEECS,ref:MultiBernoulliControl,ref:RobustMultiBernoulliControl}. Minimizing such task functions leads to selecting the action that is statistically expected to return minimum error. 
	
	Information-driven objective functions are usually information-theoretic reward functions designed to maximize the expected \textit{information gain} from the prior density to the posterior after a sensor action has been taken. Information gain is usually quantified using an information divergence from the prior to posterior. The most commonly used divergences in stochastic sensor control are R{\'e}nyi Divergence~\cite{ref:Divergences,ref:HoangControl} and CSD~\cite{ref:CauchySchwarzLMB,ref:MengMultiSensorControl,ref:SabitaCauchySchwarz}. Some objective functions combine information-driven and task-driven components. Li et al.~\cite{ref:DistributedSparsityControl} implements a sparsity-promoting objective function, which is a weighted sum of the normalized CSD and a sparsity-promoting term, that discourages multiple sensors from observing a subset of targets and ignoring the other targets. Recently a multi-sensor control objective function using the Kullback-Leibler Divergence has been proposed~\cite{ref:AidanICCAIS}.
	
	\section{Problem Statement and Background}
	\label{sec:Background}
	
	\subsection{Overall architecture}	
\begin{figure}
		\centering
		\includegraphics[width=\columnwidth]{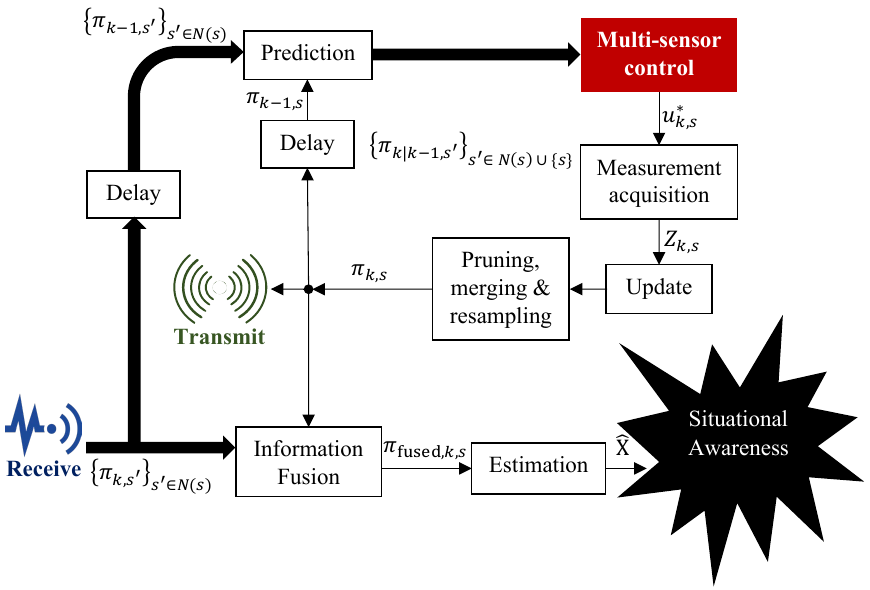}
		\caption{Overall operations running onboard each sensor node $s$, for tracking and control.}
		\label{fig:overall_blockdiagram}
	\end{figure}

	Figure~\ref{fig:overall_blockdiagram} provides a general overview of the entire sequence of operations executed by each sensor node, $s$, in a distributed sensor network of $\mathcal{S}$ sensor nodes. For every filtering time instance $k$, the node carries out the prediction step of its multi-object filter, both on its own multi-object prior, $\pi_{k-1,s}$, and on the latest multi-object posterior densities, $\pi_{k-1,s'}$, obtained at the previous filtering time $k-1$ from all adjacent sensor nodes $s'\in N^{(s)}$, where $N^{(s)}$ denotes the collection of all neighboring nodes that transceive inform with node $s$.
	
	The primary focus of this work is the design of a solution for the \textit{multi-sensor control} block, highlighted in red. This block takes as input all the predicted multi-object densities, $\pi_{k|k-1,s'}, \ s'\in N^{(s)} \cup \{s\}$, and then determines the \textit{optimal} control command (such as movement or rotation), $u_{k,s}^*$, for sensor node $s$. Following this, the system executes the control command on the sensor (e.g., moving or rotating it as required) and subsequently performs detection, yielding a measurement set $Z_{k,s}$. This set is utilized to update the predicted density, resulting in the local posterior $\pi_{k,s}$.
	
	The obtained local posterior is then transmitted to all neighboring sensors and serves as the local multi-object prior for the next time step. To achieve situational awareness regarding the number and states of objects of interest, each node fuses its local posterior with all received posteriors. It then estimates the multi-object state from the fused posterior.
	
   \subsection{System components and model assumptions} 
    The prominent component of the overall architecture shown in Figure~\ref{fig:overall_blockdiagram} is the \textit{multi-object filter} running at each node. Before the filter of choice is presented, some notation and assumptions need to be clarified.
    
    For a multi-object system presented in discrete time \(k\), the multi-object state $X_k$ can be described by a labeled random finite set (RFS) composed of several single-object states,
    \begin{equation}
    	X_k=\{(x_{1,k},\ell_1),\ldots,(x_{\mathcal{S},k},\ell_{\mathcal{S}})\}\in\mathcal{F}(\mathbb{X}\times\mathbb{L})
    \end{equation}
    where each state \(x\) has been augmented with a label \(\ell\). At each discrete filtering time step \(k\), each sensor node $s$ of the network returns a set of measurements that may include object detections and false alarms (clutter). 
    
\begin{figure}
    	\centering
    	\includegraphics[width=0.6\columnwidth]{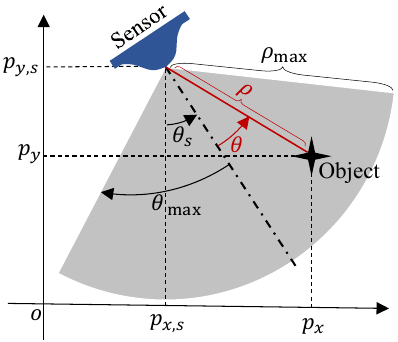}
    	\caption{Diagram of a sensor with limited FoV, and the notation used for formulation of the probability of detection.}
    	\label{fig:FoV_limits}
    \end{figure}

    The probability of detection of an object with state $x$ is denoted by \(p_{D,s,k}(x)\), implying that it may differ from one sensor node to another, may vary with time and may be dependent on the object's state. The dependence on object's state is particularly of interest and significance. The \textit{limits} of the sensor's field of view is mathematically formulated by incorporating such a dependency. An example is shown in Figure~\ref{fig:FoV_limits}. In a 2D tracking application, the sensor is looking downward at a tilt angle $\theta_s$. Its field of view is limited in range (can only detect objects at distances between $\rho_\min$ and $\rho_\max$) and angle (up to $\theta_\max$ radians away from sensor's tilted axis). Having the sensor located at $(p_{x,s},p_{y,s})$, the probability of detection of an object located at $(p_x,p_y)$ can be given by:
    \begin{equation}
            \setlength{\arraycolsep}{4pt}
    	p_{D,s}(\rho,\theta) = \left\{
    	\begin{array}{ll}
    		\frac{p_{D,\max}\tanh\left(\frac{\left[\rho_\max-\rho\right]}{\lambda}\right)}
    		{\tanh\left(\frac{\left[\rho_\max-\rho_\min\right]}{\lambda}\right)}
    		& \text{if }|\theta|\leqslant \theta_\max,\rho\leqslant\rho_\max\\
    		0 & \text{else}
    	\end{array}
    	\right.
            \setlength{\arraycolsep}{6pt}
    	\label{eq:pd_example}
    \end{equation}
    where 
    \begin{equation*}
    \rho = \sqrt{(p_x-p_{x,s})^2+(p_y-p_{y,s})^2}
    \end{equation*}
    \begin{equation*}
    \theta = \text{atan2}\left(p_x-p_{x,s},p_y-p_{y,s}\right) + \frac{\pi}{2} - \theta_s
    \end{equation*}
    where atan2 means the four-quadrant inverse tangent function.
    
\begin{figure}
    	\centering
    	\includegraphics[width=\columnwidth]{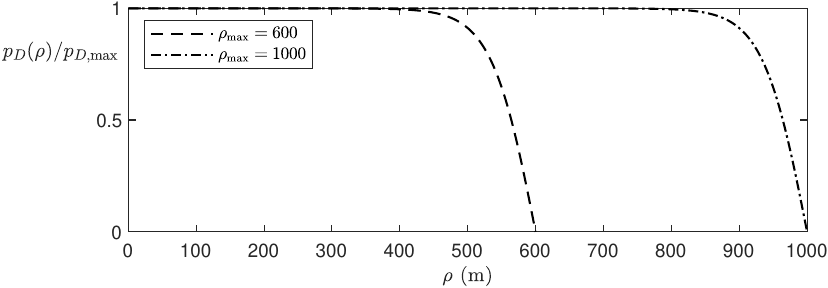}
    	\caption{Probability detection variations against range for two cases in both of which $\rho_\min = 0$ and $\lambda=65\,\text{m}$, but maximum range is 600\,m in one scenario and and 1000\,m in another}
    	\label{fig:pdplots}
    \end{figure}

    Figure~\ref{fig:pdplots} shows how the $p_{D,s}$ values given by the above model vary from the maximum value to zero as the sensor-object distance increases up to the maximum range.
    
    Conditional on detection, the sensor $s$ is assumed to provide a measurement $z$ that is distributed according to a density given by the likelihood function \(g_{s}({\cdot|x,x_s})\) where $x$ is the detected object's state, and $x_s$ denotes the sensor's state (e.g. its location $(p_{x,s},p_{y,s})$ and orientation $\theta_s$ as shown in Figure~\ref{fig:FoV_limits}). Each sensor can be controlled via a sensor control command $u^*_s\in\mathbb{U}$ where \(\mathbb{U}\) is a finite set of sensor control commands. Obviously, the sensor' s state will depend on what control action is chosen and executed. Hence, the measurement likelihood values will be dependent on sensor control actions too. A multi-sensor control command can be constructed by appending multiple single-sensor control commands, \(\mathfrak{u}=(u_1,\ldots,u_{\mathcal{S}})\in\mathbb{U}^{\mathcal{S}}\). 
    
    The overarching aim of controlling a set of sensors for multi-object tracking is to substantially improve the accuracy of target state estimation. This encompasses refining both the detection of the total number of targets (cardinality) and the detailed assessment of each target's specific attributes, such as position, velocity, and orientation.
    
    \subsection{The multi-object filter}
    As was emphasized previously, a multi-object filter runs onboard each sensor node. The Bayesian multi-object filter propagates the multi-object density through its prediction and update steps, and after the posterior is fused with other similar posteriors received from neighboring nodes, it extracts the multi-object estimate, which is our realization of situational awareness. The distributed multi-sensor control solution proposed in this paper can be applied with almost any choice of RFS-based multi-object filter. However, due to its proven performance and intuitive structure, we present and formulate our solution for applications where the labeled multi-Bernoulli (LMB) filter~\cite{reuter2015multiple} is locally running in each node. 
    
    The LMB distribution is completely described by its components \({\pi}=\{(r^{(\ell)},p^{(\ell)}(\cdot))\}_{\ell\in\mathbb{L}}\) where \(r^{(\ell)}\) is the probability of the existence of an object with label \(\ell\in\mathbb{L}\), and \(p^{(\ell)}(x)\) is the probability density of the object state conditional on its existence. Mathematically, for any multi-object state $X = \{(x_1,\ell_1),\ldots,(x_n,\ell_n)\}$, the multi-object density is given by
    \begin{equation}
        \pi(X)=\omega(\{\ell_1,\ldots,\ell_n\})\prod_{i=1}^{n} p^{(\ell_i)}(x_i)
    \end{equation}
    where
    \begin{equation}
    	\omega(L)=
    	\prod_{\ell\in L} r^{(\ell)}
    	\prod_{\ell\in (\mathbb{L} - L)}(1-r^{(\ell)})
    \end{equation}
    is the probability of joint existence of all objects with labels \(\ell \in L\) and non-existence of all other labels, if $\forall i,j\in[1,n],i\neq j \Rightarrow \ell_i\neq \ell_j$, else $\pi(X)=0$. For computing single-object densities, a particle implementation is used in this paper, where the density of each LMB component with label \(\ell\) is approximated by \(J^{(\ell)}\) weights and particles, 
    $
    	p^{(\ell)}(x)\approx\sum_{j=1}^{J^{(\ell)}}\mathrm{w}_j^{(\ell)}\delta(x-x_j^{(\ell)})
    $
    where \(\delta(\cdot)\) is the Dirac delta function. Note that the proposed multi-sensor control solution could also be used with th Gaussian-mixture implementation.
    
    \subsection{Information fusion}
    As shown in Figure~\ref{fig:overall_blockdiagram}, \textit{information fusion} is a critical task that needs to be completed in each sensor node. With limited field of view, we need a \textit{complementary} fusion method that puts together all the information gathered by various sources. We have already developed such a solution tailored for fusion of several LMB posteriors~\cite{9311857} but in a centralized sensor network. In this paper, we propose a slightly varied version that can be directly applied in a distributed sensor network.
    
    Referring to Figure~\ref{fig:overall_blockdiagram}, assume that the multiple LMB densities that need to be fused are parametrized as follows:
    \begin{align*}
    \pi_{k,s} &= \left\{\left(r_{k,s}^{(\ell)},p_{k,s}^{(\ell)}(\cdot)\right)
    \right\}_{\ell\in\mathbb{L}_{k,s}}\\
    \text{for }s' \in N^{(s)}: 
    \pi_{k,s'} &= \left\{\left(r_{k,s'}^{(\ell)},p_{k,s'}^{(\ell)}(\cdot)\right)
    \right\}_{\ell\in\mathbb{L}_{k,s'}}.
    \end{align*}
    Two important notes should be made here. Firstly, the space of object labels (detected by each sensor up to time $k$) may be different in each node due to limited FoVs. Secondly, if the single-object densities are approximated by particles, for each label, the fused single-object density will be approximated by all the particles from different sensors for that label. The particle weights will need to be re-scaled according to the probability of existence reported by each sensor for that label. 
    
    For any node $s' \in N^{(s)} \cup \{s\}$, let us assume that at time $k$, single-object density associated with label $\ell$ is approximated by $J_{k,s'}^{(\ell)}$ pair of particles and weights, where the $j$-th pair is denoted by $ 	(x_{j,k,s'}^{(\ell)},w_{j,k,s'}^{(\ell)}).$ The fused density at node $s$, $p_{\text{fused},k,s}(\cdot)$ will then be represented by $J_{\text{fused},k,s}^{(\ell)} = \sum_{s' \in N^{(s)} \cup \{s\}} J_{k,s'}^{(\ell)}$ pairs of particles and weights denoted by the union of individual sets of particles and weights, as follows:
    \begin{equation}
    	p_{\text{fused},k,s}^{(\ell)}(x)\approx\sum_{s' \in N^{(s)} \cup \{s\}} \sum_{j=1}^{J_{k,s'}^{(\ell)}} \left[\upalpha_{k,s'}^{(\ell)} w_{j,k,s'}^{(\ell)}\right]\,\delta(x-x_{j,l,s'}^{(\ell)})
    	\label{eq:fusion_p}
    \end{equation}
    where the scaling factors $\upalpha_{k,s'}^{(\ell)}$ are proportional to probabilities of existence, i.e.
    $$
    	\upalpha_{k,s'}^{(\ell)} = {r_{k,s'}^{(\ell)}}\ \Bigg/ {\sum_{\mathfrak{s}\,\in\, N^{(s)} \,\cup\,\{s\}}r_{k,\mathfrak{s}}^{(\ell)}}.
    $$

    This merging and reweighting of the particles is actually performing arithmetic averaging over the density functions~\cite{ref:AAFusionPart2, ref:AAFusionPart3}.

    \begin{figure*}[t]
    	\centering
    	\includegraphics[width=\textwidth]{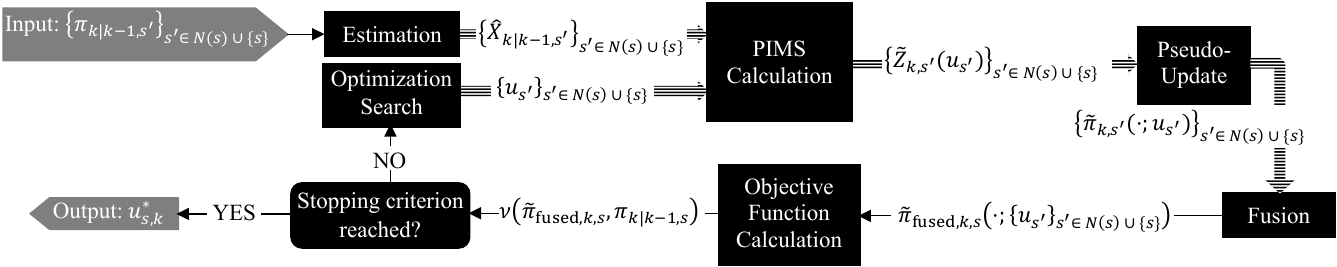}
    	\caption{The proposed architecture of the contents of multi-sensor control block executing at sensor node $s$, in the overall schematics shown in Figure~\ref{fig:overall_blockdiagram}.}
    	\label{fig:proposed_architecture}
    \end{figure*}
    
    To fuse the probabilities of existence for label $\ell$ at each node, the individual probabilities of existence are combined according to the complementary fusion rule~\cite{9311857}: 
    \begin{equation}
    	r_{\text{fused},k,s}^{(\ell)} = \left[\displaystyle\sum_{s'\in N^{(s)}\cup\{s\}}{q_{k,s'}^{(\ell)}}\right]
    	\,\Bigg/\,
    	\left[1+\sum_{s' \in N^{(s)} \cup \{s\}} q_{k,s'}^{(\ell)}\right]
    	\label{eq:fusion_r}
    \end{equation}
    where $
    	q_{k,s'}^{(\ell)} = r_{k,s'}^{(\ell)} \,\Bigg/\, \left[1 - r_{k,s'}^{(\ell)}\right].
    $
    Note that pruning and resampling the particles will ensure numerical tractability.

    \subsection{Distributed Information Fusion Challenges}
    A significant challenge in distributed information fusion implementations is label inconsistency. Label inconsistency occurs when there is a mismatch in the labels assigned to the tracked objects across different sensor nodes, manifesting as different sensors assigning the same label to multiple different objects. This overwhelmingly occurs when multiple objects are first detected by several sensors at the same time, such that through the birth process running independently in each sensor, multiple newly detected objects are assigned the same label. This arises from the conventional labeling paradigm where each object is labeled with a tuple $(k, m)$, where $k$ is the time step of birth and $m$ is an index. To resolve the issue of label inconsistency, the labels of newly detected objects take the form $(k, i, m)$, where $i$ is the identity of the sensor node that first detected the object, as in previous work~\cite{ref:KlupacsFusion}.
    
    Another challenge in distributed information fusion is double-counting, where multiple labels are applied to the same target across the network. Complementary fusion is prone to double-counting, due to any single label that has a large probability of existence $r^{(\ell)}$ will lead to a fused probability of existence $r_{\text{fused}}^{(\ell)}$ that is large. To mitigate this problem, we first ensure that the birth process of the filters are carefully devised so that clutter measurements in the middle of a sensor's FoV is rarely supported by any existing labeled densities, including birth densities. Additionally, we use a probability of existence threshold for estimation, that is set very high. Finally, when the same object is represented by more than one label, is is expected that the single-object density particles are centered very closely to each other. Therefore when the distance between the Expected A Posteriori (EAP) estimates of two objects is below a small threshold, the single-object densities are merged~\cite{ref:KlupacsFusion}.

    \section{Distributed Multi-Sensor Control: The Proposed Solution}
    \label{sec:Solution}
    In this section, we present our proposed multi-sensor control solution for distributed multi-target tracking, in a step-by-step fashion. Figure~\ref{fig:proposed_architecture} shows a block diagram of our proposed solution, in a form that could be the contents of the ``Multi-sensor Control'' block in Figure~\ref{fig:overall_blockdiagram}.
    
	\subsection{Estimation}
	The first step is to \textit{estimate} the number of objects and their states from the predicted densities at the local node $s$ and received from the neighboring nodes $s'\in N^{(s)}$. We choose the EAP estimates. In the case of LMB densities, assuming that $\pi_{k|k-1,s'}$ is parametrized as
	$
	\left\{\left(r_{k|k-1,s'}^{(\ell)}, \left\{\left(w_{j,k|k-1,s'},x_{j,k|k-1,s'}\right)\right\}_{j=1}^{J_{k-1,s'}^{(\ell)}}\right)\right\}_{\ell\in\mathbb{L}_{k-1,s'}}
	$
	the EAP estimate for number of objects (cardinality) is given by the sum of all probabilities of existence,
	\begin{equation}
		\hat{|X|}_{k|k-1,s'} = \sum_{\ell\in\mathbb{L}_{k-1,s'}} r_{k|k-1,s'}^{(\ell)}.
		\label{eq:card_EAP_estimate}
	\end{equation}
	The objects predicted to exist are then chosen as those with the highest probabilities of existence, up to $\hat{|X|}_{k|k-1,s'}$ objects. For each chosen label $\ell$, the EAP estimate for state is given by:
	\begin{equation}
		\label{eq:state_EAP_estimate}
		\hat{x}^{(\ell)}_{k|k-1,s'} = \sum_{j=1}^{J_{k-1,s'}^{(\ell)}} w_{j,k|k-1,s'}\,x_{j,k|k-1,s'}.
	\end{equation}
	\subsection{Calculation of the predicted ideal measurement set (PIMS)}
\begin{figure*}[t]
		\centering
		\begin{tabular}{ccc}
			\includegraphics[width=0.33\textwidth]{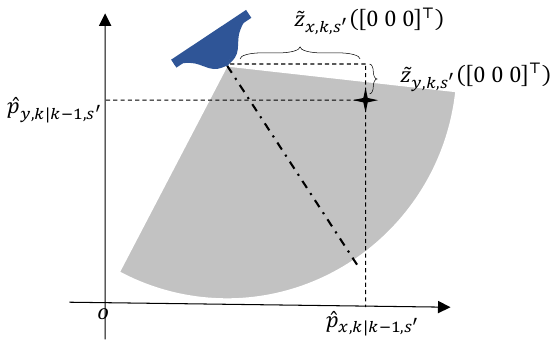}
			&
			\includegraphics[width=0.33\textwidth]{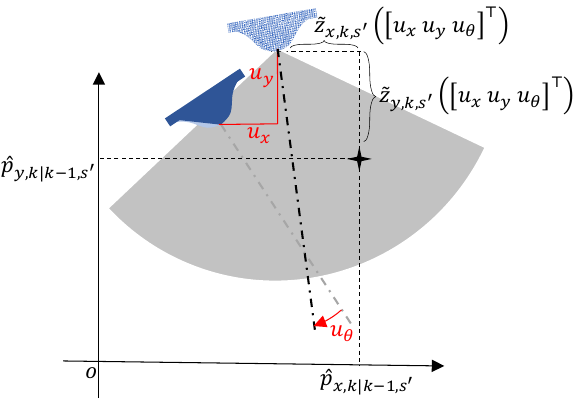}
			&
			\includegraphics[width=0.25\textwidth]{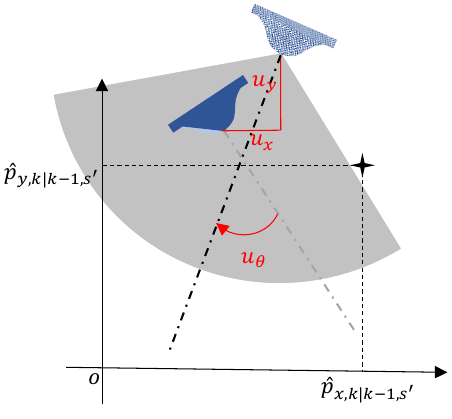}
			\\
			\footnotesize(a) & \footnotesize(b) & \footnotesize(c)
		\end{tabular}
		
		\caption{PIMS measurement returned for an object with hypothesized (a) zero control command, (b) a non-zero translation and rotation control command, and (c) another control command with the translation being same as in (b) but larger rotation that leads to the object being missed.} 
		\label{fig:PIMS_examples}
	\end{figure*}

	The next step is to compute hypothesized measurements sets that could ideally be returned by each sensor node. Such sets are \textit{ideal} in the sense that they include no measurement noise, false alarm or miss-detection, and correspond to all the objects estimated to be existing. An example is shown in Figure~\ref{fig:PIMS_examples}. Consider an object that is predicted by sensor node $s'$, to exist at an estimated location of $(\hat{p}_{x,k|k-1,s'},\hat{p}_{y,k|k-1,s'})$. As shown in Figure~\ref{fig:PIMS_examples}(a), without any sensor action (zero control command), ideally the object would be detected, returning a measurement that is comprised of the horizontal and vertical distance from the object to the sensor, denoted by $\tilde{z} = (\tilde{z}_{x,k,s'},\tilde{z}_{y,k,s'})$. If a hypothesized control command in the form of a 2D translation and rotation is applied to the sensor, the measurements will change -- see Figure~\ref{fig:PIMS_examples}(b). Due to the sensor's limited field of view, it is also possible that a predicted object is entirely missed by the sensor -- see Figure~\ref{fig:PIMS_examples}(c).
	
	\subsection{Pseudo-update and fusion}
	Once the PIMS are generated for each sensor node (self and neighboring nodes), a pseudo-update step is run. ``Pseudo'' refers to the hypothetical nature of the measurements generated with zero noise for hypothetical control actions. The results are the multi-object posteriors that would be generated locally in each sensor node if the node's locally acquired measurements were the PIMS generated in node $s$ for them.
	
	The pseudo-posteriors are then fused using the same fusion scheme that would be used in the update step after actual measurement acquisition by the sensor. In our experiments, with LMB filters running in each sensor node, fusion rules~\eqref{eq:fusion_r} and~\eqref{eq:fusion_p} are used, as also used in the update step.
	
	\subsection{Objective Function}
	
	The pseudo-posterior densities and their fused density are clearly dependent on the pseudo-measurements which are themselves dependent on the selected control commands, $\mathfrak{u}_{s} = \{u_{s'}\}_{s'\in N^{(s)}\cup\{s\}}$. The fused pseudo-posterior is input to an objective function, which is at the core of optimizing the sensing performance through sensor control. It returns a scalar value and should be formulated in such a way that its optimal value (its minimum if formulated as a cost function, or maximum if formulated as a reward function) is associated with the \textit{best statistically expected} sensing performance.

Generally, two types of objective functions have appeared in the sensor control literature: \textit{task-driven} and \textit{information-driven}. The former is usually defined to directly optimize a particular aspect of tracking, such as maximizing \textit{coverage}. Information-driven objective functions are usually information-theoretic reward functions designed to maximize the expected \textit{information gain} from the prior density to the posterior after a sensor action has occurred, usually quantified using an information divergence from the prior to the posterior.

In this work, we propose an information-theoretic objective function that has been approximated to reduce computation time, modified to penalize actions that result in targets not longer being observed. The proposed objective function is also subject to three constraints based on practicability: avoiding collisions with the targets, avoiding collisions with the other sensors, and remaining within communication distance of the other sensors. The proposed objective function is given by:

\begin{equation}
    \nu(\bm{\pi}_1,\bm{\pi}_2)=\tilde{D}_{\text{KL}}(\bm{\pi}_1, \bm{\pi}_2)+\phi(\bm{\pi}_1,\bm{\pi}_2),
    \label{eq:our_objective_fun}
\end{equation}

where $\bm{\pi}_1=\bm{\pi}_{k\vert k-1,s}(\cdot)$ and $\bm{\pi}_2=\tilde{\bm{\pi}}_{\text{fused},k,s}(\cdot;\mathfrak{u}_s)$ and the constrained optimization problem is as follows:

\begin{equation}
    \begin{array}{ll}
        \text{maximize}
        & \nu\bigg(\bm{\pi}_{k|k-1,s}(\cdot) , \tilde{\bm{\pi}}_{\text{fused},k,s}(\cdot;\mathfrak{u}_s)\bigg)
        \\
        \text{subject to} & \psi_{\tilde{\bm{\pi}}_{\text{fused},k,s}}(\epsilon(u_s))<\psi_{\text{th}}
        \\
        & \eta(\mathfrak{u}_s) > \eta_{\text{th}}
        \\
        & \zeta(\mathfrak{u}_s) > 0.
    \end{array}
    \label{eq:optimization} 
\end{equation}

Each component of the objective function and the constraints are defined and explained below.

\subsubsection{Modified Kullback-Leibler Divergence}

In the proposed objective function, the Kullback-Leibler divergence (KLD) is used as the base information-theoretic objective function. Referring to the notation from Figure~\ref{fig:proposed_architecture}, consider two multi-object densities from a node $s$ at time $k$; one being the prior $\bm{\pi}_{k|k-1,s}(\cdot)$ and the other being the fused pseudo-posterior $\tilde{\bm{\pi}}_{\text{fused},k,s}(\cdot;\mathfrak{u}_s)$ which depends on a chosen multi-sensor control action $\mathfrak{u}_s$. The sensor control objective function is the reward associated with choosing $\mathfrak{u}_s$, quantified by computing the KLD from the prior to the fused pseudo-posterior:

\begin{equation}
    \begin{array}{l}
        D_{\text{KL}}
        \bigg(
        \tilde{\bm{\pi}}_{\text{fused},k,s}(\cdot;\mathfrak{u}_s)\,||\,\bm{\pi}_{k|k-1,s}(\cdot)
        \bigg) \\
        =	
        \int_{\mathbb{X}} \,\tilde{\bm{\pi}}_{\text{fused},k,s}(X;\mathfrak{u}_s) 
        \,\log
        \left(
        \dfrac{\tilde{\bm{\pi}}_{\text{fused},k,s}(X;\mathfrak{u}_s)}{\bm{\pi}_{k|k-1,s}(X)}
        \right)\, \delta X.
    \end{array}
    \label{eq:KLD_reward}
\end{equation}
In the case that both densities are LMB, denoting them by:
$$
\tilde{\bm{\pi}}_{\text{fused},k,s} = 
    \left\{
        \left(\tilde{r}^{(\ell)}_{\text{fused},k,s},\tilde{p}^{(\ell)}_{\text{fused},k,s}\right)
    \right\}_{\ell\in\mathbb{L}_{k,s}}
$$
$$
\bm{\pi}_{k|k-1,s} = 
    \left\{
        \left(r^{(\ell)}_{k|k-1,s},p^{(\ell)}_{k|k-1,s}\right)
    \right\}_{\ell\in\mathbb{L}_{k|k-1,s}}
$$
the integral has a closed form given by~\cite{ref:LMBParticle}:
\begin{equation}
\begin{split}
    &\nu\bigg(\bm{\pi}_{k|k-1,s}(\cdot) , \tilde{\bm{\pi}}_{\text{fused},k,s}(\cdot;\mathfrak{u}_s)\bigg) \\
    &\,\,\,\,=
    \sum_{\ell\in\mathbb{L}_{k|k-1,s}}
    \Bigg[
        \tilde{r}^{(\ell)}_{\text{fused},k,s} \log\left(\frac{\tilde{r}^{(\ell)}_{\text{fused},k,s}}{r^{(\ell)}_{k|k-1,s}}\right)
        +
        \left[1-\tilde{r}^{(\ell)}_{\text{fused},k,s}\right] \times\ \\ 
        &\,\,\,\,\,\log\left(\frac{1-\tilde{r}^{(\ell)}_{\text{fused},k,s}}{1-r^{(\ell)}_{k|k-1,s}}\right) + \tilde{r}^{(\ell)}_{\text{fused},k,s} 
        D_{\text{KL}}(\tilde{p}^{(\ell)}_{\text{fused},k,s}\,||\,p^{(\ell)}_{k|k-1,s})
    \Bigg].
    \label{eq:LMB_reward}
\end{split}
\end{equation}

The KLD between two LMB densities can be approximated to reduce the computation time in the following manner. Consider an example scenario of two normal distributions over a one-dimensional state, $P$ and $Q$, with probability density functions $p=\mathcal{N}(x; \mu_p,\sigma_p)$ and $q=\mathcal{N}(x; \mu_q,\sigma_q)$. The KLD between two univariate normal distributions is given by:

\begin{equation}
    \label{eq:KLD_Normal}
    D_{KL}(P\,||\,Q)=\mathrm{log}\left(\frac{\sigma_q}{\sigma_p}\right)+\frac{\sigma_p^2+(\mu_p-\mu_q)^2}{2\sigma_q^2}-\frac{1}{2}.
\end{equation}

When the densities have flat peaks, i.e. the two variances $\sigma_p$ and $\sigma_q$ are in the same scale and are relatively large compared to the difference between the two means $\lvert\mu_p-\mu_q\rvert$, then the KLD between the two spatial distributions will be negligible compared to the other terms in~\eqref{eq:LMB_reward}, and~\eqref{eq:LMB_reward} can therefore be approximated by~\eqref{eq:approxKLD}:

\begin{equation}
\begin{split}
    \tilde{\nu}=\sum_{\ell\in\mathbb{L}_{k|k-1,s}}\left[\tilde{r}_{\text{fused},k,s}^{(\ell)}\mathrm{log}\left(\frac{\tilde{r}_{\text{fused},k,s}^{(\ell)}}{r_{k|k-1,s}^{(\ell)}}\right)+\right.\\\left.(1-\tilde{r}_{\text{fused},k,s}^{(\ell)})\times\mathrm{log}\left(\frac{1-\tilde{r}_{\text{fused},k,s}^{(\ell)}}{1-r_{k|k-1,s}^{(\ell)}}\right)\right].
    \label{eq:approxKLD}
\end{split}
\end{equation}

When the fused pseudo-update's probability of existence approaches 1, i.e. $\tilde{\nu}\approx-\mathrm{log}(r_{k|k-1,s}^{(\ell)})$, which could be very small if the prediction's probability of existence is close to 1 (i.e. the target is already observed) but could be very large if the prediction's probability of existence is close or equal to 0 (i.e. a new target is observed). When the fused pseudo-update's probability of existence approaches 0, $\tilde{\nu}\approx-\mathrm{log}(1-r_{k|k-1,s}^{(\ell)})$, which could be very large if the prediction's probability of existence is close to 1 (i.e. the target is already observed) but could be very small if the prediction's probability of existence is close or equal to 0 (i.e. no target is observed). However, if the densities have sharp peaks, then~\eqref{eq:KLD_Normal} may not be negligible compared to~\eqref{eq:approxKLD}, and including it places greater emphasis on maximizing localization accuracy.

In distributed tracking involving sensors with limited FoVs, usually \textit{maximum coverage} is the main intended outcome of multi-sensor control. Consequently, the significance of information gain through single-object densities with sharper peaks (emphasizing localization accuracy) is negligible compared to the information gain through fused probabilities of existence that convey more confidence (are closer to 0 or 1). Therefore, the last term of the reward function~\eqref{eq:LMB_reward} can be ignored to significantly speed up computation with no noticeable reduction in sensor control performance, and~\eqref{eq:approxKLD} can be used as the reward function.

Additionally, a penalty is added when a target label is included in the prior but not in the pseudo-posterior, i.e. when the actions result in a target no longer being observed. This emphasizes actions that do not result in targets being dropped. The penalty term is denoted $\phi$ and is formulated as:

\begin{equation}
    \phi=\displaystyle\sum_{\ell\in[\mathbb{L}_{k\vert k-1,s}-\mathbb{L}_{\mathrm{fused},k,s}]}\mathrm{log}(1-r_{k\vert k-1,s}^{(\ell)}).
    \label{eq:penalty_term}
\end{equation}
Note that this penalty term negates the second term in~\eqref{eq:LMB_reward},
$
\left[1 - \tilde{r}^{(\ell)}_{\text{fused},k,s}\right]\log\left((1-r_{k|k-1,s}^{(\ell)})\big/(1-\tilde{r}^{(\ell)}_{\text{fused},k,s})\right)
$,
when $\tilde{r}^{(\ell)}_{\text{fused},k,s}$ equals zero.

The approximated KLD between the prediction and fused pseudo-posterior LMB densities, shown in~\eqref{eq:approxKLD}, with the additional penalty term $\phi$~\eqref{eq:penalty_term}, results in the proposed objective function~\eqref{eq:our_objective_fun}.

\subsubsection{Constraints}

In practical multi-object tracking scenarios, it is important that sensors avoid collisions with both the targets that they are tracking and the other sensors in the network. Additionally, it is important that sensors do not take actions that would take them out of communication range of the rest of the network. The objective function can account for these additional motives in the following ways:

Firstly, we add a constraint on actions that would move the sensor too close to the targets. The void probability functional (or void probability) of a point process is the probability that a given region does not contain any points~\cite{ref:VoidProbability}. In the multi-object tracking problem, it is the probability that a given region does not contain any targets. This can be used to prioritize sensor actions that do not move the sensor in a way that targets enter an exclusion area around the sensor. This exclusion area $\epsilon(u_s)$, a circular region centered on the sensor with radius $\rho_\epsilon$, depends on the action $u_s$ taken by the sensor (the single-sensor control command element of $\mathfrak{u}_s$ that controls sensor $s$). The void probability of a SMC-LMB density~\cite{ref:ConstrainedCSDControl} is defined as:
\begin{multline}
    \psi_{\tilde{\bm{\pi}}_{\text{fused},k,s}}\left(\epsilon(u_s)\right)=\\\displaystyle\prod_{\ell\in\mathbb{L}_{\text{fused},k,s}}\left(1-\tilde{r}_{\text{fused},k,s}^{(\ell)}\sum_{j=1}^{J_{k,s}^{(\ell)}}1_{\epsilon(u_s)}(\tilde{x}_{\text{fused},j,k,s}^{(\ell)})\tilde{w}_{\text{fused},j,k,s}^{(\ell)}\right)
    \label{eq:VoidProbability}
\end{multline}
where
\begin{equation}
    1_{\epsilon(u)}(\tilde{x}_{\text{fused},j,k,s}^{(\ell)})=\left\{
    \begin{array}{ll}
        1
        & \text{if }\tilde{x}_{\text{fused},j,k,s}^{(\ell)}\text{ is in region }\epsilon(u_s)
        \\
        0 & \text{else}.
    \end{array}
    \right.
\label{eq:Indicator}
\end{equation}

Because $\sum_{j=1}^{J_k^{(\ell)}}w_{\text{fused},j,k,s}^{(\ell)}=1$, for a particular label, if all the particles for a particular label are inside the exclusion region, then the term in~\eqref{eq:VoidProbability} becomes $(1-\tilde{r}_{\text{fused},k,s}^{(\ell)})$. If all the particles are outside the exclusion region, it becomes $1$.

To avoid collision with other sensors, we also define a sensor sparsity term $\eta(\mathfrak{u}_s)$ that is equal to the minimum distance between sensor $s$ and a different sensor in the network after actions $\mathfrak{u}_s$ are taken:

\begin{equation}
    \eta(\mathfrak{u}_s)=\min_{s'\in \mathcal{S}\setminus s}d(s,s').
    \label{eq:collision}
\end{equation}

    where

    \begin{equation}
        d(s,s')=\sqrt{(p_{x,s}-p_{x,s'})^2+(p_{y,s}-p_{y,s'})^2}
    \end{equation}

Finally, we add a constraint on actions that would move the sensor out of communication range with the rest of the network. Clearly, this is an undesirable action that would degrade the tracking accuracy of the network, e.g. if a target is only observed by the now disconnected sensor, the rest of the network will lose this target. This is accomplished by determining if the position of sensor $s$ after actions $\mathfrak{u}_s$ is closer than $d_{\text{th}}$ to at least one other sensor in the network, where the distance threshold $d_{\text{th}}$ is the maximum communication range of all other sensors in the network, with a small buffer to account for variability and noise in sensor movements and communication range in practical scenarios,

\begin{equation}
        \zeta(\mathfrak{u}_s)=\left\{
        \begin{array}{ll}
            1
            & \text{if }\exists s'\in \mathcal{S}\setminus s,d(s,s')<d_{\text{th}}
            \\
            0 & \text{else}.
        \end{array}
        \right.
        \label{eq:communication}
    \end{equation}

The values from~\eqref{eq:VoidProbability},~\eqref{eq:collision}, and~\eqref{eq:communication} are used in the constraints in~\eqref{eq:optimization}.
	
	\subsection{Optimization and stopping criterion}
	
	A different choice of the multi-sensor control command, $\mathfrak{u}_s = \{u_{s'} \in \mathbb{U}\}_{s'\in N^{(s)}\cup \{s\}}$, would generate a different set of fused pseudo-updated probabilities of existence, and therefore a different value for the reward function~\eqref{eq:our_objective_fun}. The ultimate aim is to find the optimal multi-sensor control command at each sensor node that would return \textit{maximum} reward. Denoting the optimal command by $\mathfrak{u}_s^* = \{u_{s'}^*\}_{s'\in N^{(s)}\cup \{s\}}$, the algorithm only outputs the one element $u_s^*$ at each node $s$ which is then executed to move the sensor before actual measurement acquisition.
 
    The process of determining the solution for the multi-sensor control problem is a combinatorial optimization search, requiring an iterative optimization algorithm to avoid significant computational cost. The recently proposed distributed coordinate descent method~\cite{ref:AidanICCAIS}, to significantly reduce computational cost, only considers a sensor's neighborhood when calculating the multi-sensor control command, rather than the entire sensor network. There are non-trivial barriers to implementing a distributed control scheme where each sensor node accounts for every node's actions, including ensuring distributed consensus and computational limitations.
	
	We propose a flooding-based method that realizes a truly distributed, asynchronous, multi-sensor control algorithm. The basic idea is that each sensor node will iteratively update its own selected action, taking into account the actions that every other sensor has selected. In a distributed network, two sensors cannot know the other's selected action in real-time. But they can \textit{flood-in} the information related to the action most recently selected by others, and \textit{flood-out} their own selection to others for future use in their own asynchronous time. 
	
	At each sensor node, an iterative operation occurs with each iteration comprising flooding-in, optimization, and flooding-out steps. Let us denote the iteration counter for this process by $t$, in contrast to filtering iterations denoted by $k$. In iteration $t$, the sensor node $s$ first receives all optimal decisions $u^*_{s'}(t-1)$ from its neighboring sensors $s' \in N^{(s)}$. That is the realization of the flood-in step. Note that communication of only the selected control commands from other nodes normally does not incur a heavy load on communication. Hence, we can practically assume that after a number of flooding steps, dependent on the network size, each sensor node will have communicated with the entire network, or in the case of an extremely large network a large subset of the entire network. Denoting the total number of nodes in the network by $\mathcal{S}$ and using integer indices for sensor labels, e.g. $\mathbb{S} = \{1,\ldots,\mathcal{S}\}$, the optimization step of iteration $t$ involves an exhaustive search over the space of single-sensor commands as follows:
	\begin{equation}
		\label{eq:flood_in_optimization}
		u_s^*(t) = \underset{\bm{u}\in\mathbb{U}}{\text{arg\,max}}\ \  \nu\left(u^*_{1:(s-1)}(t-1),\bm{u},u^*_{(s+1):\mathcal{S}}(t-1)\right)
	\end{equation}
	where $u^*_{1:(s-1)}(t-1)$ denotes the $(s-1)$-tuple of recent decisions made by sensor nodes $1:(s-1)$, and similar definition applies to $u^*_{(s+1):\mathcal{S}}(t-1)$ i.e. 
	\begin{equation}
    	\begin{array}{rcl}
    		u^*_{1:(s-1)}(t-1) & \triangleq & \bigg(u^*_1(t-1),\ldots,u^*_{s-1}(t-1)\bigg) \\
    		u^*_{(s+1):\mathcal{S}}(t-1) & \triangleq & \bigg(u^*_{s+1}(t-1),\ldots,u^*_{\mathcal{S}}(t-1)\bigg).
    	\end{array}
            \label{eq:flooding_update}
	\end{equation}
	In the final step of iteration $t$ at node $s$, the locally decided optimal command for the sensor is flooded-out to all the other sensor nodes for use in their own optimization step.
	
\begin{figure*}
		\centering
		\includegraphics[width=0.95\textwidth]{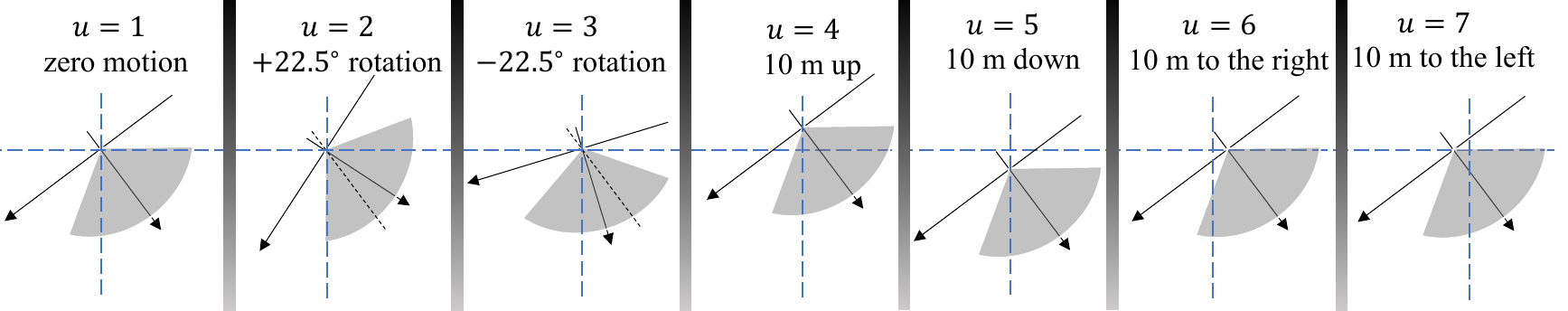}
		\caption{Possible sensor commands in the challenging case of our numerical experiments.}
		\label{fig:sensor_commands}
	\end{figure*}

	An important note to make is that solving the optimization problem in~\eqref{eq:flood_in_optimization} by exhaustive search is not computationally prohibitive. The main reason is that it involves computing the reward for just the finite number of possible single-sensor commands. In our numerical experiments, each sensor can execute one of the seven possible translation and rotation movements depicted in Figure~\ref{fig:sensor_commands}. In addition, note that when computing the reward for each possible sensor command, the PIMS and pseudo update calculations only need to be performed for the actual node $s$ and not for the others.
	
    \textbf{Stopping criterion:} The proposed method has a well defined stopping criterion, where convergence of the multi-sensor control command is mathematically guaranteed when it occurs. Let $\mathfrak{u}^*_s(t)$ denote the multi-sensor command decision made at node $s$ in iteration $t$. As it was explained above, we have:
	\begin{equation}
		\mathfrak{u}^*_s(t) = \begin{bmatrix}
			u^*_1(t-1) & \ldots & u^*_{s}(t) & \ldots & u^*_{\mathcal{S}}(t-1)
		\end{bmatrix}
		\label{eq:flooding_iterated_multi_sensor_command}
	\end{equation}
    As the optimization algorithm iterates, two possible results will occur: either $\mathfrak{u}_s^*(t)$ will converge and will not change in any future iterations, or $\mathfrak{u}_s^*(t)$ will enter a ``cycle'', where as the iterations continue, $\mathfrak{u}_s^*(t)$ will follow a repeating pattern. The iterative process should be stopped when either of these results occur, which is when the following condition is met:
    \begin{multline}
        \text{If }\exists\,s\in\mathbb{S},1<t'<t\\ \text{ s.t. }[\mathfrak{u}_s^*(t-1)=\mathfrak{u}_s^*(t'-1)]\land[\mathfrak{u}_s^*(t)=\mathfrak{u}_s^*(t')]
        \label{eq:flooding_stopping_criterion}
    \end{multline}
    The stopping condition described by~\eqref{eq:flooding_stopping_criterion} is reached when there exists a sensor $s$, and a previous iteration $t'$ that is greater than 1 and less than the current iteration $t$, such that the following condition is true: the selected multi-sensor control command for sensor $s$ at iteration $t'-1$ is the same as the selected multi-sensor control command at sensor $s$ at iteration $t-1$, and the selected multi-sensor control command for sensor $s$ at iteration $t'$ is the same as the selected multi-sensor control command at sensor $s$ at iteration $t$.

    It can be shown that either when $\mathfrak{u}_s^*(t)$ has converged (if $t' = t-1$) or when $\mathfrak{u}_s^*(t)$ has become trapped in a ``cycle'' (if $t'<t-1$), that all nodes in the network will have reached the same selected multi-sensor control command. The following lemma proves that when the criterion~\eqref{eq:flooding_stopping_criterion} is met, for all nodes in the network, $\{s'\in\mathbb{S}\}$, that $\mathfrak{u}_s^*(t)=\mathfrak{u}_{s'}^*(t)$.
 
    \begin{lemma}
        \label{lemma:convergence}
        $\text{If }\exists\,s\in\mathbb{S},1<t'<t\text{ such that }[\mathfrak{u}_s^*(t-1)=\mathfrak{u}_s^*(t'-1)]\land[\mathfrak{u}_s^*(t)=\mathfrak{u}_s^*(t')]\text{ then }\forall\,s'\in\mathbb{S},\mathfrak{u}_{s'}^*(t)=\mathfrak{u}_{s'}^*(t')$.
    \end{lemma}

    \noindent\textit{\textbf{Proof by contradiction.}}

    $\text{If } \exists\, s\in\mathbb{S},1<t'<t\text{ such that }[\mathfrak{u}_s^*(t-1)=\mathfrak{u}_s^*(t'-1)]\land[\mathfrak{u}_s^*(t)=\mathfrak{u}_s^*(t')]\text{ then }$
\begin{equation}
    \exists\, s'\in\mathbb{S}\text{ such that }\mathfrak{u}_{s'}^*(t)\neq\mathfrak{u}_{s'}^*(t').
    \label{eq:contradiction}
\end{equation}

 For this statement to be true, there must exist an element of $\mathfrak{u}_{s'}^*(t)$ that is not equal to the equivalent element of $\mathfrak{u}_{s'}^*(t')$, i.e.
 \begin{multline}
     \exists\, n\in\mathbb{S}\text{ such that the n-th element of } \mathfrak{u}_{s'}^*(t)\neq\\ \text{ the n-th element of }\mathfrak{u}_{s'}^*(t').
     \label{eq:n-th_element}
 \end{multline}
 There are two distinct possibilities for $n$:
 \begin{itemize}
    \item[--] $n\neq s'$: In this case, it can be seen from~\eqref{eq:flooding_update} that the n-th element of $\mathfrak{u}_{s}^*(t)$ is equal to $u_n^*(t-1)$ and the n-th element of $\mathfrak{u}_{s}^*(t')$ is equal to $u_n^*(t'-1)$. If~\eqref{eq:n-th_element} is true however, then $u_n^*(t-1)\neq u_n^*(t'-1)$, therefore $\mathfrak{u}_s^*(t)\neq\mathfrak{u}_s^*(t')$, which is contradictory to~\eqref{eq:contradiction}.
    \item[--] $n=s'$: In this case, it can be seen from~\eqref{eq:flood_in_optimization} that:
    \begin{multline*}
        u_{s'}^*(t) =\\ \underset{\bm{u}\in\mathbb{U}}{\text{arg\,max}}\  \nu\,\bigg(  u_{1:(s'-1)}^*(t-1),\bm{u},u_{(s'+1):\mathcal{S}}^*(t-1)\bigg)
    \end{multline*} 
    \begin{multline}
        u_{s'}^*(t') =\\ \ \ \ \ \ \ \ \ \ \underset{\bm{u}\in\mathbb{U}}{\text{arg\,max}}\  \nu\,\bigg( u_{1:(s'-1)}^*(t'-1),\bm{u},u_{(s'+1):\mathcal{S}}^*(t'-1)\bigg).
        \label{eq:two_argmaxes}
    \end{multline} 
    
    If any $\exists\, n\in\mathbb{S}\setminus s'$ where $u_n^*(t-1)\neq u_n^*(t'-1)$, then the previous case ($n\neq s'$) is true. If not, then the input arguments of the two reward functions in~\eqref{eq:two_argmaxes} are equal, and consequently the search outcomes, $u_{s'}^*(t)$ and $u_{s'}^*(t')$ must equal as well, which is contradictory to~\eqref{eq:contradiction}. \hfill \qedsymbol
\end{itemize}

In practice, once the convergence required by the stopping criterion~\eqref{eq:flooding_stopping_criterion} is met at some node $s$, in the next iteration of information flooding, that sensor will broadcast a signal containing $\mathfrak{u}_s^*(t)$ to the other sensors, telling them that convergence is reached and to stop iterating and execute their received optimal control command.

Another practical consideration is related to the asynchronous nature of this algorithm. In practice, sensors in the center of the network require fewer flooding iterations to receive information from all communicable sensors. Thus, in each single iteration, while different nodes should take similar amounts of time to reach their newly selected control command, small variations are possible, causing some sensors to progress to the next iteration earlier. Therefore, each sensor $s$ will have an individual iteration $t_s$. However, as the values of multi-sensor commands at each sensor come from other sensors at their own iterations via flooding, the fastest sensor cannot progress more than one iteration beyond the slowest sensor. In other words, at any time, $t_{\text{fastest node}} < t_{\text{slowest node}}+1\text{ filtering iteration}$.

\section{Numerical Experiments}
\label{sec:Experiment}
\subsection{Scenarios}
\begin{figure}[ht]
   \centering
   \includegraphics[width=0.8\columnwidth]{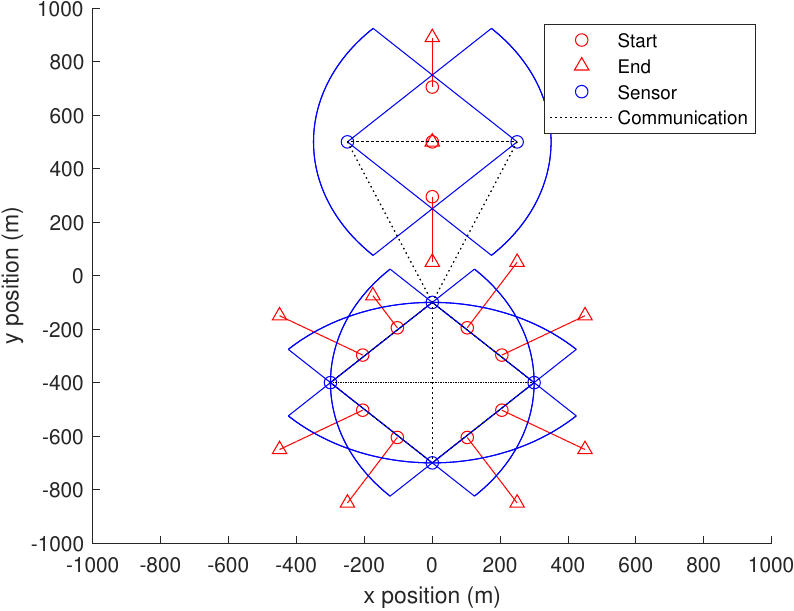}
   \caption{Overview of the 6 sensor scenario: Initial sensor positions are represented by blue circles, the initial FoV of each sensor is shown by the blue lines, the presence of communication between a pair of sensors is indicated by a dotted line, and the target trajectories are denoted in red.}
   \label{fig:6sensor_scenario}
\end{figure}

Multiple challenging multi-object tracking scenario were developed to evaluate the performance of the DF-SC method. Figure~\ref{fig:6sensor_scenario} illustrates the first scenario, with 6 sensors tracking 11 targets for a total of 50 time steps. The targets are distributed within a $800\,\text{m}\times 2000$\,m area and follow linear trajectories. All of the targets are present at the first time step and are observed by at least one of the sensors; however, as the scenario progresses, the targets will move away from the center of the scenario and therefore out of the initial sensor FoVs. This dynamic nature of the multi-target tracking problem is handled by the complementary information fusion between sensors, so that targets aren't dropped when they move out of a sensor's FoV but remain in other sensors' FoVs and they aren't double-counted when multiple sensors observe the same target. Additionally, two of the targets will die prematurely.

The detection profile of each sensor is modeled by~\eqref{eq:pd_example}, in the form depicted in Figure~\ref{fig:FoV_limits}, with $\theta_\max = 45^\circ$ (90$^\circ$ field of view) and a maximum detection range of $\rho_\max = 600$\,m. The communication range between sensors is simulated to be at maximum 800\,m. Each sensor has 3 possible actions: remain stationary, rotate $22.5^\circ$ clockwise, and rotate $22.5^\circ$ anticlockwise.

\begin{figure}[ht]
   \centering
   \includegraphics[width=0.8\columnwidth]{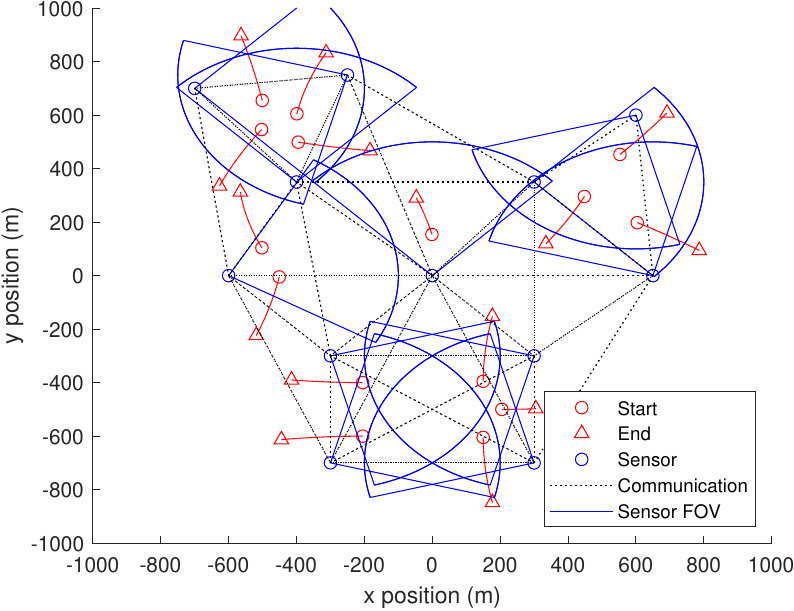}
   \caption{Overview of the 12 sensor scenario.}
   \label{fig:12sensor_scenario}
\end{figure}

Figure~\ref{fig:12sensor_scenario} illustrates the second scenario, with 12 sensors tracking 15 targets for a total of 50 time steps. This scenario includes both target births during the first 25 time steps and target deaths during the last 25 time steps. In the scenario the targets follow a noisy constant turn motion mode and the maximum detection range of the sensors is set to 500\,m, while the maximum communication range between sensors is kept at 800\,m. In the second scenario, each sensor has 7 possible actions: remain stationary, rotate $22.5^\circ$ clockwise, rotate $22.5^\circ$ anticlockwise, move 10\,m up, move 10\,m left, move 10\,m down, and move 10\,m right.

\subsection{Methods}
The proposed Distributed Flooding method (DF-SC) for multi-sensor control is evaluated and compared against several other approaches. Sensors with fixed positions and orientations are examined to establish baseline results. An individual sensor control (I-SC) method where each sensor independently finds the optimal action for itself without considering the other sensor's actions is evaluated to investigate the benefit of performing cooperative multi-sensor control. For the I-SC approach, the fused pseudo-updated posterior in the objective function~\eqref{eq:our_objective_fun} is replaced by the local pseudo-updated posterior, and each sensor independently performs an exhaustive search over the possible single-sensor control commands.

Finally, the recently proposed Distributed Coordinate Descent approach (DCD-SC) is also tested, to determine the comparative multi-object tracking accuracy and computation time. Wang et al.~\cite{ref:Xiaoying} describes the relationship between the number of coordinate descent runs required to reach a likelihood of finding the global optimum. Given $\mathfrak{M}$ local optima and the probability of finding the global optimum $P_{\mathrm{success}}$, the number of required runs is given by~\eqref{eq:coordinate_descent}.

\begin{equation}
    m=\left\lceil\frac{\mathrm{log}(1-P_{\mathrm{success}})}{\mathrm{log}(1-\frac{1}{\mathfrak{M}})}\right\rceil
    \label{eq:coordinate_descent}
\end{equation}

Assuming that $\mathfrak{M}=2\lvert N^{(s)}\rvert$, then to achieve a $95\%$ probability of finding the global optimum in the first scenario, the central sensor with the most neighbors requires 29 runs, and in the second scenario, the central sensor with the most neighbors requires 53 runs. This is an impractical number of runs due to the excessive computational cost, therefore, we decided to aim for a sub-optimal but computationally tractable solution. To compare the computation time and accuracy trade-off, several different numbers of runs are tested.

\subsection{Results}

\begin{figure}[ht]
   \centering
   \includegraphics[width=0.8\columnwidth]{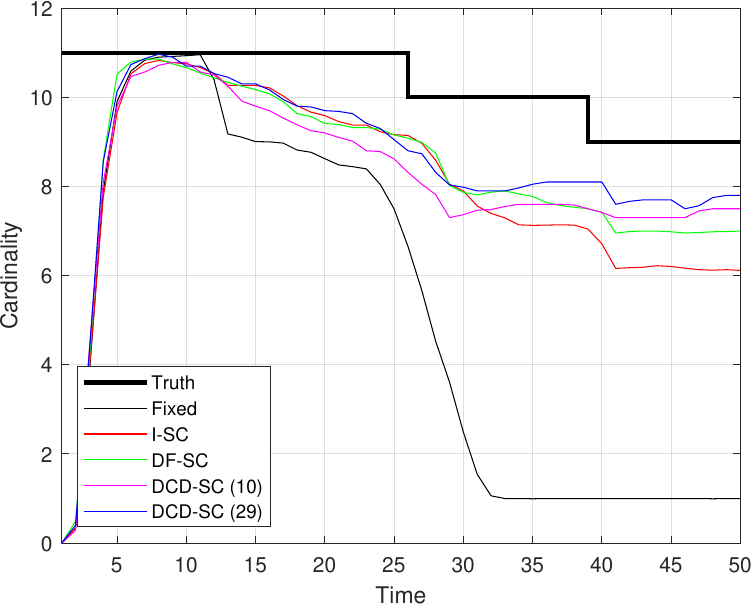}
   \caption{The average cardinality error of the fixed sensors (top), I-SC (middle), and DF-SC (bottom) methods. The true number of targets at each time step is indicated by the thin black line, and the average estimated number of targets is shown by the colored circles.}
   \label{fig:average_cardinality_error}
\end{figure}

Figure~\ref{fig:average_cardinality_error} shows the average estimated target cardinality (number of estimated targets) across the 30 Monte Carlo runs at each time step, for each of the four tested methods. The closer the thin lines (estimated target cardinality) are to the thick black line (true target cardinality), the more accurate the number of estimated targets and therefore multi-object tracking accuracy.

\begin{table}[ht]
    \centering
    \caption{Average multi-object tracking performance results over 30 Monte Carlo runs of fixed sensors, I-SC, DCD-SC, and DF-SC.}
    \begin{tabular}{l|rrr}
    	\small
        \small\textbf{Control method} & \small\textbf{OSPA\,(m)} & \small\textbf{OSPA$^{(2)}$\,(m)} & \small\textbf{Comp. time\,(s)} \\
        \hline
        \small Fixed Sensors & \small 56.40 & \small 54.60 & \small N/A \\
        \small I-SC & \small 26.38 & \small 35.22 & \small 0.21 \\
        \small DCD-SC (1 run) & \small 26.10 & \small 42.14 & \small 2.33 \\
        \small DCD-SC (5 runs) & \small 24.49 & \small 34.05 & \small 7.24 \\
        \small DCD-SC (10 runs) & \small 23.86 & \small 34.98 & \small 10.40 \\
        \small DCD-SC (29 runs) & \small 20.26 & \small 32.77 & \small 16.32 \\
        \small \textbf{DF-SC} & \small \textbf{22.90} & \small \textbf{33.19} & \small \textbf{2.82} \\
    \end{tabular}
    \label{tab:6_sensor_results}
\end{table}

Table~\ref{tab:6_sensor_results} details the average OSPA and OSPA$^{(2)}$ distances across the 30 Monte Carlo runs for each method, as well as the average computation time per sensor for each method.

For the second scenario, due to containing a larger number of both sensors and targets, and the results from the first scenario showing the high computational cost required for the DCD-SC method to achieve similar performance to the proposed DF-SC method, DCD-SC was omitted.

\begin{figure}[ht]
   \centering
   \includegraphics[width=0.8\columnwidth]{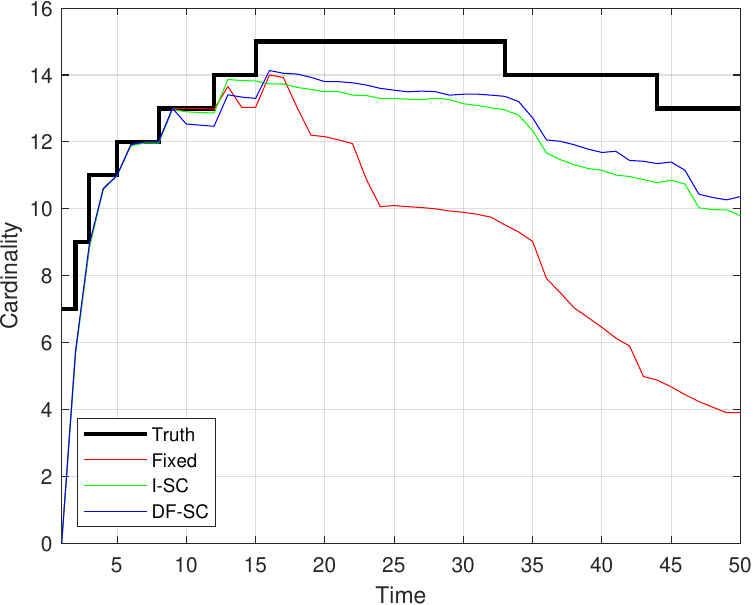}
   \caption{The average cardinality error of the fixed sensors (top), I-SC (middle), and DF-SC (bottom) methods. The true number of targets at each time step is indicated by the thin black line, and the average estimated number of targets is shown by the colored circles.}
   \label{fig:12cams_average_cardinality_error}
\end{figure}

Figure~\ref{fig:12cams_average_cardinality_error} shows the average estimated target cardinality (number of estimated targets) across the 30 Monte Carlo runs at each time step, for each of the tested methods. The closer the thin lines (estimated target cardinality) are to the thick black line (true target cardinality), the more accurate the number of estimated targets and therefore multi-object tracking accuracy.

\begin{table}[ht]
    \centering
    \caption{Average multi-object tracking performance results over 30 Monte Carlo runs of fixed sensors, I-SC, and DF-SC.}
    \begin{tabular}{l|rrr}
    	\small
        \small\textbf{Control method} & \small\textbf{OSPA\,(m)} & \small\textbf{OSPA$^{(2)}$\,(m)} & \small\textbf{Comp. time\,(s)} \\
        \hline
        \small Fixed Sensors & \small 35.58 & \small 32.51 & \small N/A \\
        \small I-SC & \small 17.68 & \small 18.87 & \small 0.30 \\
        \small \textbf{DF-SC} & \small \textbf{16.31} & \small \textbf{22.18} & \small \textbf{10.89} \\
    \end{tabular}
    \label{tab:12_sensor_results}
\end{table}

Table~\ref{tab:12_sensor_results} details the average OSPA and OSPA$^{(2)}$ distances across the 30 Monte Carlo runs for each method, as well as the average computation time per sensor for each method.

\subsection{Discussion}

As discussed previously, the main source of multi-object tracking error in a scenario with a large number of mobile, dynamic targets, is the targets not being observed by any of the sensors, i.e. cardinality error. For the first scenario, Figure~\ref{fig:average_cardinality_error} shows how when the sensors are fixed, while initially the targets are all within the FoV of the sensor network, when the targets begin to move outside of the network's FoV the targets are rapidly lost, leading to a high cardinality error and poor multi-object tracking performance. In comparison, when the various multi-sensor control methods are used, the sensors are able to follow the trajectories of the targets, leading to a much lower cardinality error. The proposed DF-SC method has a lower cardinality error than I-SC, indicating that the cooperative nature of DF-SC leads to more complete coverage than the individual nature of I-SC, improving multi-object tracking performance. DF-SC has a lower average cardinality error than DCD-SC with 10 runs, however, it has a higher average cardinality error than DCD-SC with 29 runs.

Table~\ref{tab:6_sensor_results} highlights the significant improvement that multi-sensor control has over fixed, stationary sensors when tracking multiple dynamic targets, as well as the significant reduction in both the OSPA and OSPA$^{(2)}$ distances in DF-SC compared to I-SC, although at the cost of a significantly longer computation time (\textapprox 13 times longer).

Figure~\ref{fig:12cams_average_cardinality_error} again shows that the proposed DF-SC has significantly less cardinality error than both fixed cameras and I-SC, further indicating that the cooperative nature of DF-SC allows for better coverage than individual sensor control in complex multi-target tracking scenarios. Table~\ref{tab:12_sensor_results} highlights that DF-SC has the lowest OSPA error of all the methods tested, although I-SC has a lower OSPA$^{(2)}$, this was due to some track inconsistency in the DF-SC simulations. The computation time for DF-SC is also significantly longer in comparison to I-SC compared to the scenario in Figure~\ref{fig:6sensor_scenario}, \textapprox 36 times longer.

The following is a comparison of the computational costs associated with each tested method. The computational complexity of I-SC is $\mathcal{O}\left(\lvert\mathbb{U}\rvert\right)$. The computational complexity of DCD-SC is  $\mathcal{O}\left(m \times n \times \lvert\mathbb{U}\rvert\right)$, where $m$ represents the number of runs, $n$ is the number of iterations per run, and $\lvert\mathbb{U}\rvert$ is the number of possible control commands. The computational complexity of DF-SC is expressed as $\mathcal{O}\left(t \times \lvert\mathbb{U}\rvert\right)$, with $t$ being the number of flooding iterations and $\lvert\mathbb{U}\rvert$ representing the number of possible control commands.

On average, for the same scenario, DCD-SC is computationally $\frac{m \times n}{t}$ times heavier than DF-SC. Under the assumption that the number of coordinate descent iterations to convergence will tend to be similar to the number of flooding iterations to convergence, $n\approx t$, then DCD-SC is approximately $m$ times computationally heavier than DF-SC. One practical consideration is that each evaluation of a potential multi-sensor control command in DF-SC is slower than in DCD-SC, as it evaluates the actions of $\mathcal{S}$ sensors rather than $N^{(s)}$, therefore in larger, sparsely connected networks, the computational cost advantage that DF-SC enjoys over DCD-SC may diminish. Interestingly, neither method's computational cost depends on the number of possible actions, $\lvert U\rvert$.
    
    \section{Conclusion}
    \label{sec:Conclusion}
    This paper addresses the problem of multi-sensor control in a distributed sensor network with a focus on multi-target tracking applications. An information-theoretic objective function with collision-avoidance and network connectivity constraints was proposed. A distributed control method based on distributed flooding (DF-SC) was proposed and subsequently compared to the Distributed Coordinate Descent (DCD-SC) method as well as individual sensor control (I-SC). To evaluate the tracking performance of these methods, two scenarios with varying sensor and target numbers were designed, and the OSPA distance and OSPA$^{(2)}$ error metrics were employed.

    The results demonstrated that the DF-SC method outperformed the fixed sensors and I-SC method and matched or outperformed DCD-SC in terms of accuracy in the first scenario. With a higher number of iterations, the performance of DCD-SC was better than DF-SC, however this came at a significant computational cost. In the second, more complex, scenario, again DF-SC outperformed the fixed sensors, and achieved a lower cardinality error than I-SC, demonstrating the advantages of performing cooperative multi-sensor control in multi-object tracking.

    Upon comparing the computational complexity of the tested methods, it was determined that DF-SC is significantly computationally advantageous over DCD-SC for comparable performances, especially when there are fewer sensors in the network and when the sensor network is more densely connected. As a direction for future research, adaptive methods that dynamically adjust the control parameters based on tracking performance and network density could be explored.
    
    \section*{Acknowledgment}

The Australian Research Council supported this work through Grant DE210101181.
    \bibliographystyle{IEEEtran}
    \bibliography{AidanBib.bib}

\end{document}